\documentclass[prd,nofootinbib,showpacs,preprint]{revtex4}
\usepackage{amsmath,amssymb}
\usepackage{epsfig}
\usepackage{graphicx}
\usepackage[usenames,dvipsnames]{color}
\usepackage{slashed}
\usepackage{slashed}
\usepackage{amsmath}
\usepackage{pdfpages}
\usepackage{color}

\begin{document}
\title{Galactic Center Gamma Ray Excess in a \\Radiative Neutrino Mass Model}
\author{Debasish Borah}
\email{dborah@tezu.ernet.in}
\affiliation{Department of Physics, Tezpur University, Tezpur - 784028, India}
\author{Arnab Dasgupta}
\email{arnabdasgupta28@gmail.com}
\affiliation{Centre for Theoretical Physics, Jamia Millia Islamia - Central University, Jamia Nagar, New Delhi - 110025, India}

\begin{abstract}
The Fermi gamma ray space telescope data have pointed towards an excess of gamma rays with a peak around $1-3$ GeV in the region surrounding the galactic center. This anomalous excess can be described well by a dark matter candidate having mass in the range $31-40$ GeV annihilating into $b\bar{b}$ pairs with a cross section of $\langle \sigma v \rangle \simeq (1.4-2.0) \times 10^{-26} \; \text{cm}^3/\text{s}$. In this work we explore the possibility of having such a dark matter candidate within the framework of a radiative neutrino mass model. The model is a simple extension of the standard model by an additional $U(1)_X$ gauge symmetry where the standard model neutrino masses arise both at tree level as well as radiatively by the anomaly free addition of one singlet fermion $N_R$ and two triplet fermions $\Sigma_{1R}, \Sigma_{2R}$ with suitable Higgs scalars. The spontaneous gauge symmetry breaking is achieved in such a way which results in a residual $Z_2$ symmetry and hence providing a stable cold dark matter candidate. We show that the singlet fermionic dark matter candidate in our model can give rise to the galactic center gamma ray excess. The parameter space which simultaneously satisfy the constraints on relic density, direct detection scattering as well as collider bounds essentially corresponds to an s-wave resonance where the gauge boson mass $m_X$ is approximately twice that of dark matter mass $m_{\chi}$. We also discuss the compatibility of such a light fermion singlet dark matter with light neutrino mass.
\end{abstract}
\maketitle

\section{Introduction}
\label{sec:intro}
Recent analysis of Fermi Gamma Ray Space Telescope data has shown an excess of gamma ray from the Galactic Center (GC) with a feature similar to annihilating dark matter \cite{GCfeb} (For a review of dark matter, please see \cite{Jungman:1995df}). Previous studies \cite{GCprev} also identified a similar excess of $1-3$ GeV gamma rays from the region surrounding the GC. According to the analysis presented in \cite{GCfeb}, the Fermi telescope signal of gamma ray excess in the GC can very well be fit by a $31-40$ GeV dark matter particle annihilating into $b\bar{b}$ pairs with an annihilation cross section of $\sigma v = (1.4-2.0) \times 10^{-26} \; \text{cm}^3/s$, normalized to a local dark matter density of $0.3 \; \text{GeV}/\text{cm}^3$. The required annihilation cross section is coincidentally very close to the annihilation cross section of typical Weakly Interacting Massive Particle (WIMP) dark matter candidate in order to produce the correct dark matter relic abundance observed by the Planck experiment \cite{Planck13} 
\begin{equation}
\Omega_{\text{DM}} h^2 = 0.1187 \pm 0.0017
\label{dm_relic}
\end{equation}
where $\Omega$ is the density parameter and $h = \text{(Hubble Parameter)}/100$ is a parameter of order unity.

Several interesting particle physics models have already been proposed \cite{Berlin:2014tja, GCmodels} which explain the GC excess of gamma rays. Here we study the possibility of providing such an explanation within the framework of an abelian extension of standard model, originally proposed by \cite{Adhikari:2008uc} and later studied in the context of dark matter and eV scale sterile neutrino in \cite{Borah:2012qr} and \cite{Borah:2014} respectively. The salient feature of the model is the way it relates dark matter with neutrino mass where neutrino masses arise at one loop level with dark matter particles running inside the loops: more popularly known as "scotogenic" model \cite{Ma:2006km}. The additional abelian gauge symmetry $U(1)_X$ and the corresponding gauge charges for the fields are chosen in such a way that it gives rise to a remnant $Z_2$ symmetry so that the lightest $Z_2$-odd particle is stable and hence can be a cold dark matter candidate. As studied in details in \cite{Borah:2012qr}, this model has several dark matter candidates namely, fermion singlet, fermion triplet, scalar singlet and scalar doublet. Scalar dark matter phenomenology is similar to the Higgs portal models discussed extensively in the literature. In these scenarios, the scalar dark matter annihilates into the Standard Model (SM) particles through the Higgs boson. Co-annihilations through gauge bosons can also play a role if the CP even and CP odd components of the neutral Higgs have a tiny mass difference as discussed recently in \cite{arnabborah}. In the context of GC gamma ray excess, several Higgs portal models have already been studied and there exists at least one neutral Higgs lighter than the SM Higgs which acts as a mediator between scalar dark matter and the SM particles. The mass of this light neutral Higgs is approximately equal to twice the scalar dark matter mass in order to satisfy experimental bounds on relic density as well as direct detection experiments.

Instead of pursuing Higgs portal like scalar dark matter scenarios in the model, we study the fermionic dark matter sector. Since the neutral component of fermion triplet needs to be very heavy ($ 2.28-2.42 \; \text{TeV} $) in order to reproduce correct dark matter relic density \cite{Ma:2008cu}, we confine our discussion to fermion singlet dark matter in this work. That is, we explore the possibility of fermion singlet dark matter in this model with mass around 30 GeV which can simultaneously give rise to GC gamma ray excess as well satisfy dark matter bounds on relic density as well as direct detection cross section. Such a light fermion singlet dark matter particle will self-annihilate through the abelian vector boson $X$ into SM particles. We also incorporate the collider constraints on such additional vector boson and its gauge couplings. We find that, although the relic density and direct detection constraints allow a significant region of the parameter space, the collider constraints reduce the parameter space into the s-wave resonance region where the gauge boson mass is approximately twice that of dark matter mass. Finally, we check whether such a light fermion singlet dark matter is compatible with neutrino mass which arise at one loop level.

This letter is organized as follows: in section \ref{model}, we briefly discuss the model. In section \ref{sec:darkmatter}, we discuss the singlet fermion dark matter as a source of GC gamma ray excess taking into account all necessary experimental constraints. In section \ref{neutrino}, we discuss the compatibility of light singlet fermion dark matter with neutrino mass and finally conclude in section \ref{conclude}. 

\section{The Model}
\label{model}
The model which we take as a starting point of our discussion was first proposed in \cite{Adhikari:2008uc}. The authors in that paper discussed various possible scenarios with different combinations of Majorana singlet fermions $N_R$ and Majorana triplet fermions $\Sigma_R$. Here we discuss one of such models which we find the most interesting for our purposes. This, so called model C by the authors in \cite{Adhikari:2008uc}, has the the following particle content shown in table \ref{table1}.

\begin{center}
\begin{table}
\caption{Particle Content of the Model}
\begin{tabular}{|c|c|c|c|}
\hline
Particle & $SU(3)_c \times SU(2)_L \times U(1)_Y$ & $U(1)_X$ & $Z_2$ \\
\hline
$ (u,d)_L $ & $(3,2,\frac{1}{6})$ & $n_1$ & + \\
$ u_R $ & $(\bar{3},1,\frac{2}{3})$ & $\frac{1}{4}(7 n_1 -3 n_4)$ & + \\
$ d_R $ & $(\bar{3},1,-\frac{1}{3})$ & $\frac{1}{4} (n_1 +3 n_4)$ & +\\
$ (\nu, e)_L $ & $(1,2,-\frac{1}{2})$ & $n_4$ & + \\
$e_R$ & $(1,1,-1)$ & $\frac{1}{4} (-9 n_1 +5 n_4)$ & + \\
\hline
$N_R$ & $(1,1,0)$ & $\frac{3}{8}(3n_1+n_4)$ & - \\
$\Sigma_{1R,2R} $ & $(1,3,0)$ & $\frac{3}{8}(3n_1+n_4)$ & - \\
$ S_{1R}$ & $(1,1,0)$ & $\frac{1}{4}(3n_1+n_4)$ & + \\
$ S_{2R}$ & $(1,1,0)$ & $-\frac{5}{8}(3n_1+n_4)$ & - \\
\hline
$ (\phi^+,\phi^0)_1 $ & $(1,2,-\frac{1}{2})$ & $\frac{3}{4}(n_1-n_4)$ & + \\
$ (\phi^+,\phi^0)_2 $ & $(1,2,-\frac{1}{2})$& $\frac{1}{4}(9n_1-n_4)$ & + \\
$(\phi^+,\phi^0)_3 $ & $(1,2,-\frac{1}{2})$& $\frac{1}{8}(9n_1-5n_4)$ & - \\
\hline
$ \chi_1 $ & $(1,1,0)$ & $-\frac{1}{2}(3n_1+n_4)$ & + \\
$ \chi_2 $ & $(1,1,0)$ & $-\frac{1}{4}(3n_1+n_4)$ & + \\
$ \chi_3 $ & $(1,1,0)$ & $-\frac{3}{8}(3n_1+n_4)$ & - \\
$ \chi_4 $ & $(1,1,0)$ & $-\frac{3}{4}(3n_1+n_4)$ & + \\
\hline
\end{tabular}
\label{table1}
\end{table}
\end{center}
The third column in table \ref{table1} shows the $U(1)_X$ quantum numbers of various fields which satisfy the anomaly matching conditions. The Higgs content chosen above is not arbitrary and is needed, which leads to the possibility of radiative neutrino masses in a manner proposed in \cite{Ma:2006km} as well as a remnant $Z_2$ symmetry. Two more singlets $S_{1R}, S_{2R}$ are required to be present to satisfy the anomaly matching conditions. In this model, the quarks couple to $\Phi_1$ and charged leptons to $\Phi_2$ whereas $(\nu, e)_L$ couples to $N_R, \Sigma_R$ through $\Phi_3$ and to $S_{1R}$ through $\Phi_1$. The extra four singlet scalars $\chi$ are needed to make sure that all the particles in the model acquire mass. The lagrangian which can be constructed from the above particle content has an automatic $Z_2$ symmetry and hence provides a cold dark matter candidate in terms of the lightest odd particle under this $Z_2$ symmetry. Part of the scalar potential of this model relevant for our future discussion can be written as
$$ V_s \supset \mu_1 \chi_1 \chi_2 \chi^{\dagger}_4 + \mu_2 \chi^2_2 \chi^{\dagger}_1 +\mu_3 \chi^2_3 \chi^{\dagger}_4 + \mu_4 \chi_1 \Phi^{\dagger}_1 \Phi_2 + \mu_5 \chi_3 \Phi^{\dagger}_3 \Phi_2 +\lambda_{13} (\Phi^{\dagger}_1 \Phi_1)(\Phi^{\dagger}_3 \Phi_3)$$
$$ +f_1\chi_1\chi^{\dagger}_2\chi^2_3+f_2\chi^3_2\chi^{\dagger}_4+f_3 \chi_1 \chi^{\dagger}_3\Phi^{\dagger}_1\Phi_3 +f_4 \chi^2_2\Phi^{\dagger}_1\Phi_2 + f_5 \chi^{\dagger}_3\chi_4 \Phi^{\dagger}_3 \Phi_2 $$
\begin{equation}
 +\lambda_{23} (\Phi^{\dagger}_2 \Phi_2)(\Phi^{\dagger}_3 \Phi_3)+ \lambda_{16} (\Phi^{\dagger}_1 \Phi_1)(\chi^{\dagger}_3 \chi_3) + \lambda_{26} (\Phi^{\dagger}_2 \Phi_2)(\chi^{\dagger}_3 \chi_3)
\label{scalpot}
\end{equation}

Let us denote the vacuum expectation values (vev) of various Higgs fields as $ \langle \phi^0_{1,2} \rangle = v_{1,2}, \; \langle \chi^0_{1,2,4} \rangle  =u_{1,2,4}$. We also denote the coupling constants of $SU(2)_L, U(1)_Y, U(1)_X$ as $g_2, g_1, g_X$ respectively. The charged weak bosons acquire mass $M^2_W = \frac{g^2_2}{2}(v^2_1+v^2_2) $. The neutral gauge boson masses in the $(W^{\mu}_3, Y^{\mu}, X^{\mu})$ basis is 
\begin{equation}
M =\frac{1}{2}
\left(\begin{array}{cccc}
\ g^2_2(v^2_1+v^2_2) & g_1g_2(v^2_1+v^2_2) &  M^2_{WX} \\
\ g_1g_2(v^2_1+v^2_2) & g^2_1(v^2_1+v^2_2) & M^2_{YX} \\
\ M^2_{WX} & M^2_{YX}  & M^2_{XX}
\end{array}\right)
\end{equation}
where 
$$M^2_{WX} = -g_2g_X(\frac{3}{4}(n_1-n_4)v^2_1+\frac{1}{4}(9n_1-n_4)v^2_2) $$
$$ M^2_{YX} = -g_1g_X(\frac{3}{4}(n_1-n_4)v^2_1+\frac{1}{4}(9n_1-n_4)v^2_2)$$
$$ M^2_{XX} = g^2_X(\frac{9}{4}(n_1-n_4)^2v^2_1+\frac{1}{4}(9n_1-n_4)^2v^2_2+\frac{1}{16}(3n_1+n_4)^2(4u^2_1+u^2_2+9u^2_4)) $$
The mixing between the electroweak gauge bosons and the additional $U(1)_X$ boson as evident from the above mass matrix should be very tiny so as to be in agreement with electroweak precision measurements. The stringent constraint on mixing can be avoided by assuming a very simplified framework where there is no mixing between the electroweak gauge bosons and the extra $U(1)_X$ boson. Therefore $ M^2_{WX} = M^2_{YX} = 0$ which gives rise to the following constraint
\begin{equation}
3(n_4-n_1)v^2_1 = (9n_1-n_4)v^2_2
\label{zeromixeq}
\end{equation}
which implies $1 < n_4/n_1 <9 $. If $U(1)_X$ boson is observed at LHC this ratio $n_4/n_1$ could be found empirically 
from its decay to $q\bar{q}$, $l\bar{l}$ and $\nu\bar{\nu}$ \cite{Adhikari:2008uc}. Here, $q$, $l$ and $\nu$ correspond to 
quarks, charged leptons and neutrinos respectively.
In terms of the charged weak boson mass, we have 
$$ v^2_1 = \frac{M^2_W(9n_1-n_4)}{g^2_2(3n_1+n_4)}, \quad v^2_2 = \frac{M^2_W(-3n_1+3n_4)}{g^2_2(3n_1+n_4)} $$
Assuming zero mixing, the neutral gauge bosons of the Standard Model have masses
$$ M_B = 0, \quad M^2_Z = \frac{(g^2_1+g^2_2)M^2_W}{g^2_2} $$
which corresponds to the photon and weak Z boson respectively. The $U(1)_X$ gauge boson mass is 
\begin{equation}
 M^2_X = 2g^2_X (-\frac{3M^2_W}{8g^2_2}(9n_1-n_4)(n_1-n_4)+\frac{1}{16}(3n_1+n_4)^2(4u^2_1+u^2_2+9u^2_4))
\label{mxmass}
\end{equation}

\section{Singlet Fermion Dark Matter}
\label{sec:darkmatter}
The relic abundance of a dark matter particle $\chi$ is given by the Boltzmann equation
\begin{equation}
\frac{dn_{\chi}}{dt}+3Hn_{\chi} = -\langle \sigma v \rangle (n^2_{\chi} -(n^{eqb}_{\chi})^2)
\end{equation}
where $n_{\chi}$ is the number density of the dark matter particle $\chi$ and $n^{eqb}_{\chi}$ is the number density when $\chi$ was in thermal equilibrium. $H$ is the Hubble expansion rate of the Universe and $ \langle \sigma v \rangle $ is the thermally averaged annihilation cross section of the dark matter particle $\chi$. In terms of partial wave expansion $ \langle \sigma v \rangle = a +b v^2$. Numerical solution of the Boltzmann equation above gives \cite{Kolb:1990vq}
\begin{equation}
\Omega_{\chi} h^2 \approx \frac{1.04 \times 10^9 x_F}{M_{Pl} \sqrt{g_*} (a+3b/x_F)}
\label{eq:relic}
\end{equation}
where $x_F = m_{\chi}/T_F$, $T_F$ is the freeze-out temperature, $g_*$ is the number of relativistic degrees of freedom at the time of freeze-out. Dark matter particles with electroweak scale mass and couplings freeze out at temperatures approximately in the range $x_F \approx 20-30$. More generally, $x_F$ can be calculated from the relation 
\begin{equation}
x_F = \ln \frac{0.038gm_{PL}m_{\chi}<\sigma v>}{g_*^{1/2}x_f^{1/2}}
\label{xf}
\end{equation}
where $g$ is the number of internal degrees of freedom of the dark matter particle $\chi$. The thermal averaged annihilation cross section $\langle \sigma v \rangle$ is given by \cite{Gondolo:1990dk}
\begin{equation}
\langle \sigma v \rangle = \frac{1}{8m^4T K^2_2(m/T)} \int^{\infty}_{4m^2}\sigma (s-4m^2)\surd{s}K_1(\surd{s}/T) ds
\end{equation}
where $K_i$'s are modified Bessel functions of order $i$, $m$ is the mass of Dark Matter particle and $T$ is the temperature.

The singlet Majorana fermion $N_R$ can be a dark matter candidate if it is the lightest among the $Z_2$-odd particles in the model. To calculate the relic density of $N_R$, we need to find out  its annihilation cross-section to standard model particles. For zero $Z-X$ mixing, the dominant annihilation channel is the one with $X$ boson mediation. Since the singlet fermion $N_R$ is of Majorana type, it has only axial coupling to the vector boson. The annihilation cross-section of $N_R$ into SM fermion anti-fermion pairs $f\bar{f}$ through s-channel $X$ boson \cite{Berlin:2014tja} can be written as
\begin{align}
\sigma &= \frac{n_c}{12\pi s\left[(s-m^2_{X})^2+M^2_X\Gamma^2_X\right]}\bigg{[}\frac{1-4m^2_f/s}{1-4M^2_X/s}\bigg{]}^{1/2}\times \nonumber \\
&\bigg{[}g^2_{fa}g^2_{\chi a}\bigg{(}4m^2_{\chi}\bigg{[}m^2_f\bigg{(}7-\frac{6s}{M^2_X}+\frac{3s^2}{M^4_X}\bigg{)}-s\bigg{]}+s(s-4m^2_f)\bigg{)} \nonumber \\
&+g^2_{fv}g^2_{\chi a}(s+2m^2_f)(s-4m^2_{\chi})\bigg{]}
\end{align}
Expanding in powers of $v^2$ gives 
$\sigma v $ in the form $a + b v^2$ where $a$ and $b$ are given by
\begin{widetext}
\begin{align}
a &= \frac{n_c g^2_{fa}m^2_fg^2_{\chi a}m^2_{\chi}}{24\pi^2m^2_{\chi}((M^2_X - 4 m^2_{\chi})^2+M^2_X\Gamma^2_X)}\sqrt{1-\frac{m^2_f}{m^2_{\chi}}}\bigg{(}-36 + 48 \frac{m^2_{\chi}}{m^2_f}-96\frac{m^2_{\chi}}{M^2_X}+192\frac{m^4_{\chi}}{M^4_X}\bigg{)} \nonumber \\
b &= a\bigg{[}-\frac{1}{4}+\frac{2m^2_{\chi}(M^2_X-4m^2_\chi)}{(M^2_X-4m^2_\chi)^2+M^2_X\Gamma^2_X}+\frac{1}{8(m^2_\chi-m^2_f)m^2_f}\nonumber \\
&+\frac{\left(-16+2\frac{g^2_{fv}}{g^2_{fa}}+28\frac{m^2_\chi}{m^2_f}+4\frac{g^2_{fv}m^2_\chi}{g^2_{fa}m^2_f}-24\frac{m^2_\chi}{M^2_X}+96\frac{m^4_\chi}{M^4_X}\right)}{\left(-36 + 48 \frac{m^2_{\chi}}{m^2_f}-96\frac{m^2_{\chi}}{M^2_X}+192\frac{m^4_{\chi}}{M^4_X}\right)}\bigg{]}
\end{align}
\end{widetext}

The Decay width of the $X$ boson denoted by $\Gamma_X$ is given by
\begin{align}
\Gamma_{X\rightarrow \chi\overline{\chi}} &= \frac{n_cM_Xg^2_X}{12\pi S}\bigg{[}1-\frac{4m^2_{\chi}}{m^2_{X}}\bigg{]}^{3/2} \nonumber \\
\Gamma_{X \rightarrow f \overline{f}} &= \sum_{f} \frac{n_cM_X}{12\pi S}\bigg{[}1-\frac{4m^2_f}{M^2_X}\bigg{]}^{1/2}\bigg{[}g^2_{fa}\bigg{(}1-\frac{4m^2_f}{m^2_{X}}\bigg{)} \nonumber \\ 
&+g^2_{fv}\bigg{(}1+2\frac{m^2_f}{M^2_X}\bigg{)}\bigg{]}
\end{align}
The mass of the gauge boson $X$ in the above expressions is given by equation (\ref{mxmass}). For simplicity, we assume $u_1 = u_2 = u_4 = u$ such that the mass of $X$ boson can be written as
\begin{align}
M^2_X &= 2g^2_X\bigg{[}-3\frac{m^2_W}{8g^2_2}(9n_1-n_4)(n_1-n_4)+\frac{7}{8}(3n_1+n_4)^2u^2\bigg{]}
\end{align}
The couplings $g_{fv}, g_{fa}, g_{\chi v}, g_{\chi a}$ of fermions and dark matter to $X$ boson are tabulated in the table \ref{table:coupling}.
\begin{table}[!h]
\label{table:coupling}
\centering
\begin{tabular}{|c|c|c|c|}
\hline
& $n_c$ & $g_{fv}/g_X$ & $g_{fa}/g_X$ \\
\hline \hline
$l=e,\mu,\tau$ & 1 & $\frac{9}{8}\left(n_4-n_1\right)$ & $\frac{1}{8}\left(n_4-9n_1\right)$   \\
$\nu_l$ & 1 & $\frac{n_4}{2}$ & $-\frac{n_4}{2}$ \\
$U=u,c$ & 3 & $\frac{1}{8}(11n_1-n_4)$& $\frac{3}{8}(n_1-n_4)$ \\
$D=d,s,b$ & 3 & $\frac{1}{8}(5n_1+3n_4)$ & $\frac{3}{9}(n_4-n_1)$ \\
$N_R$ & 1 & 0 & $\frac{3}{8}(3n_1+n_4)$ \\
\hline
\end{tabular}
\caption{Couplings of SM particles and dark matter to the vector boson $X$}
\end{table}

\begin{figure*}[!h] 
\centering
\begin{tabular}{c}
\epsfig{file=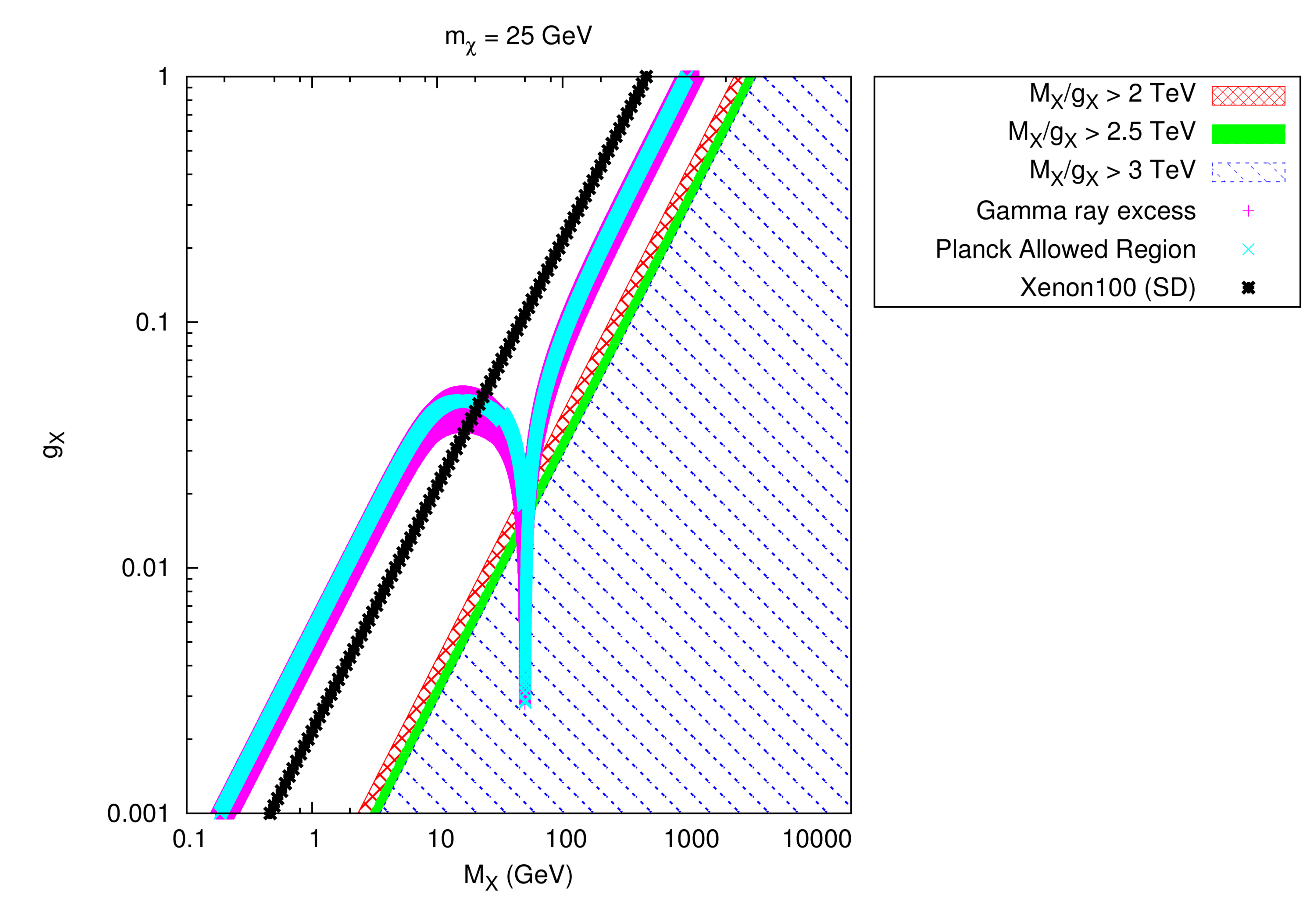,width=1.0\linewidth,clip=}\\
\end{tabular}
\caption{Parameter space in the $g_X-M_X$ plane for dark matter mass $m_{\chi} = 25$ GeV. The red-hatched, green and blue dot-dashed regions correspond to the allowed region after the constraints on $M_X/g_X$ are imposed. The area to the left of the black line is ruled out by Xenon100 bounds on direct detection cross section. The solid blue and pink regions correspond to regions favored by the relic density and galactic center excess respectively.}
\label{fig1}
\end{figure*}

\begin{figure*}[!h] 
\centering
\begin{tabular}{c}
\epsfig{file=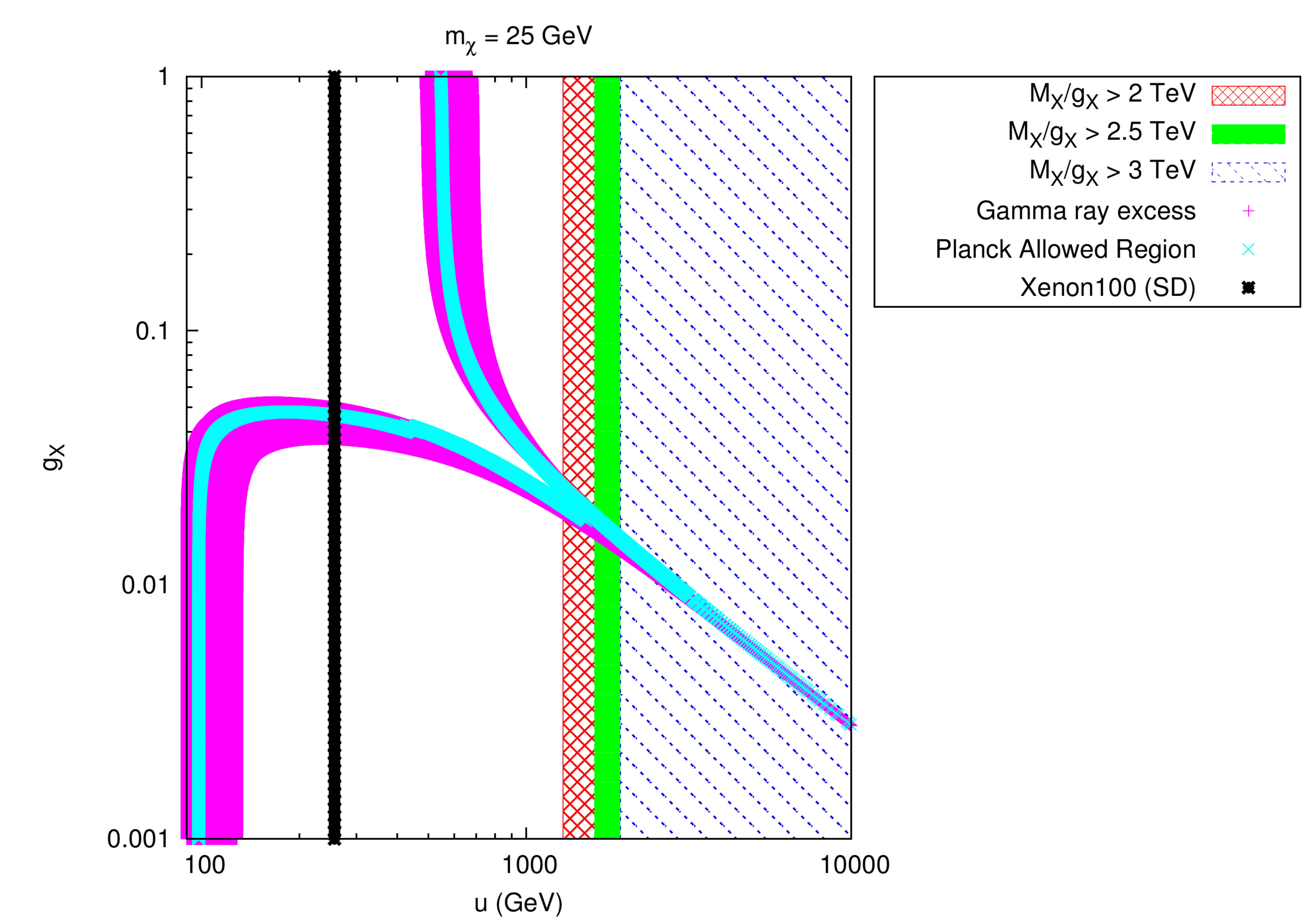,width=1.0\linewidth,clip=}\\
\end{tabular}
\caption{Parameter space in the $g_X-u$ plane for dark matter mass $m_{\chi} = 25$ GeV. The red-hatched, green and blue dot-dashed regions correspond to the allowed region after the constraints on $M_X/g_X$ are imposed. The area to the left of the black line is ruled out by Xenon100 bounds on direct detection cross section. The solid blue and pink regions correspond to regions favored by the relic density and galactic center excess respectively.}
\label{fig2}
\end{figure*}

\begin{figure*}[!h] 
\centering
\begin{tabular}{c}
\epsfig{file=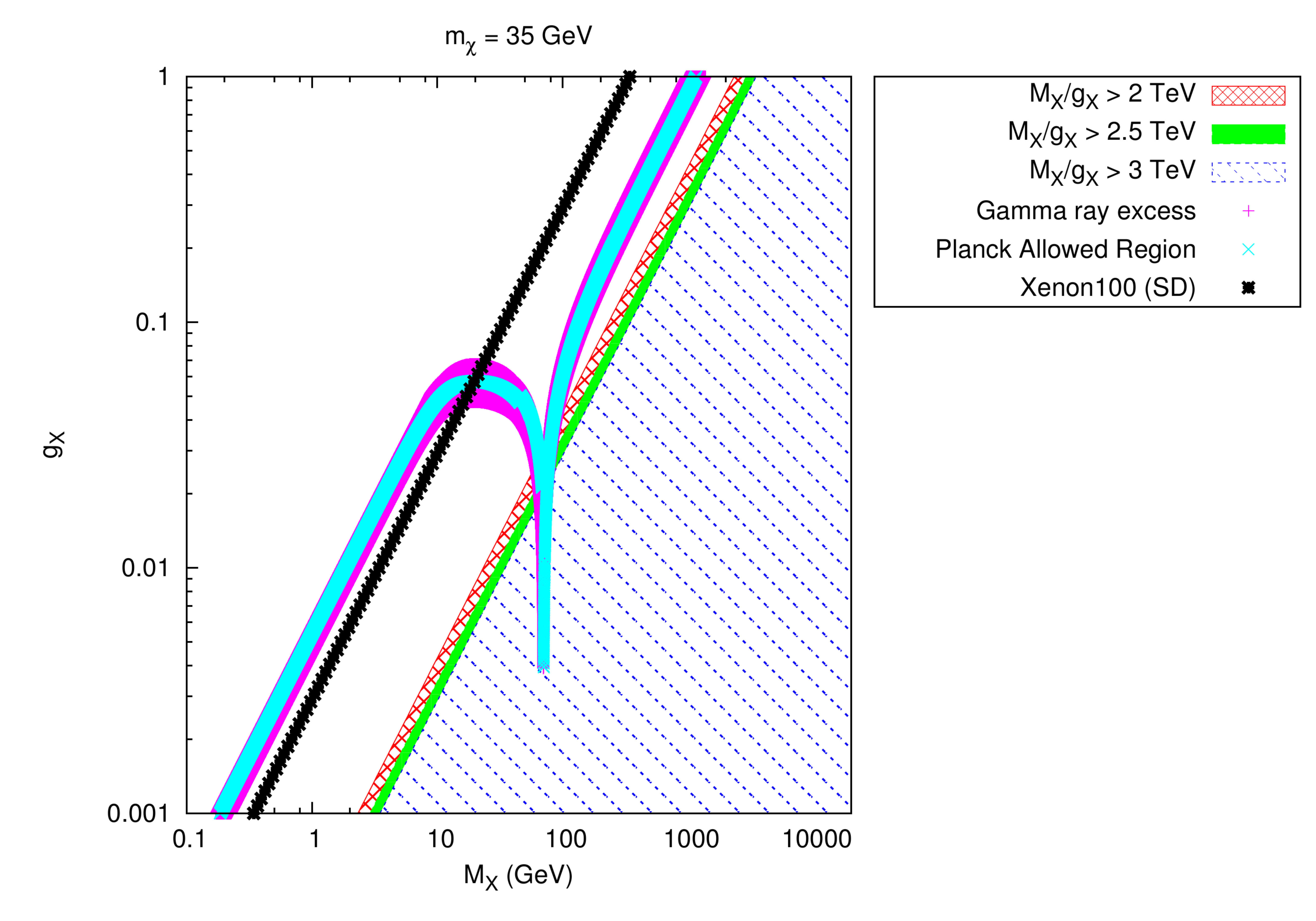,width=1.0\linewidth,clip=}\\
\end{tabular}
\caption{Parameter space in the $g_X-M_X$ plane for dark matter mass $m_{\chi} = 35$ GeV. The red-hatched, green and blue dot-dashed regions correspond to the allowed region after the constraints on $M_X/g_X$ are imposed. The area to the left of the black line is ruled out by Xenon100 bounds on direct detection cross section. The solid blue and pink regions correspond to regions favored by the relic density and galactic center excess respectively.}
\label{fig3}
\end{figure*}

\begin{figure*}[!h] 
\centering
\begin{tabular}{c}
\epsfig{file=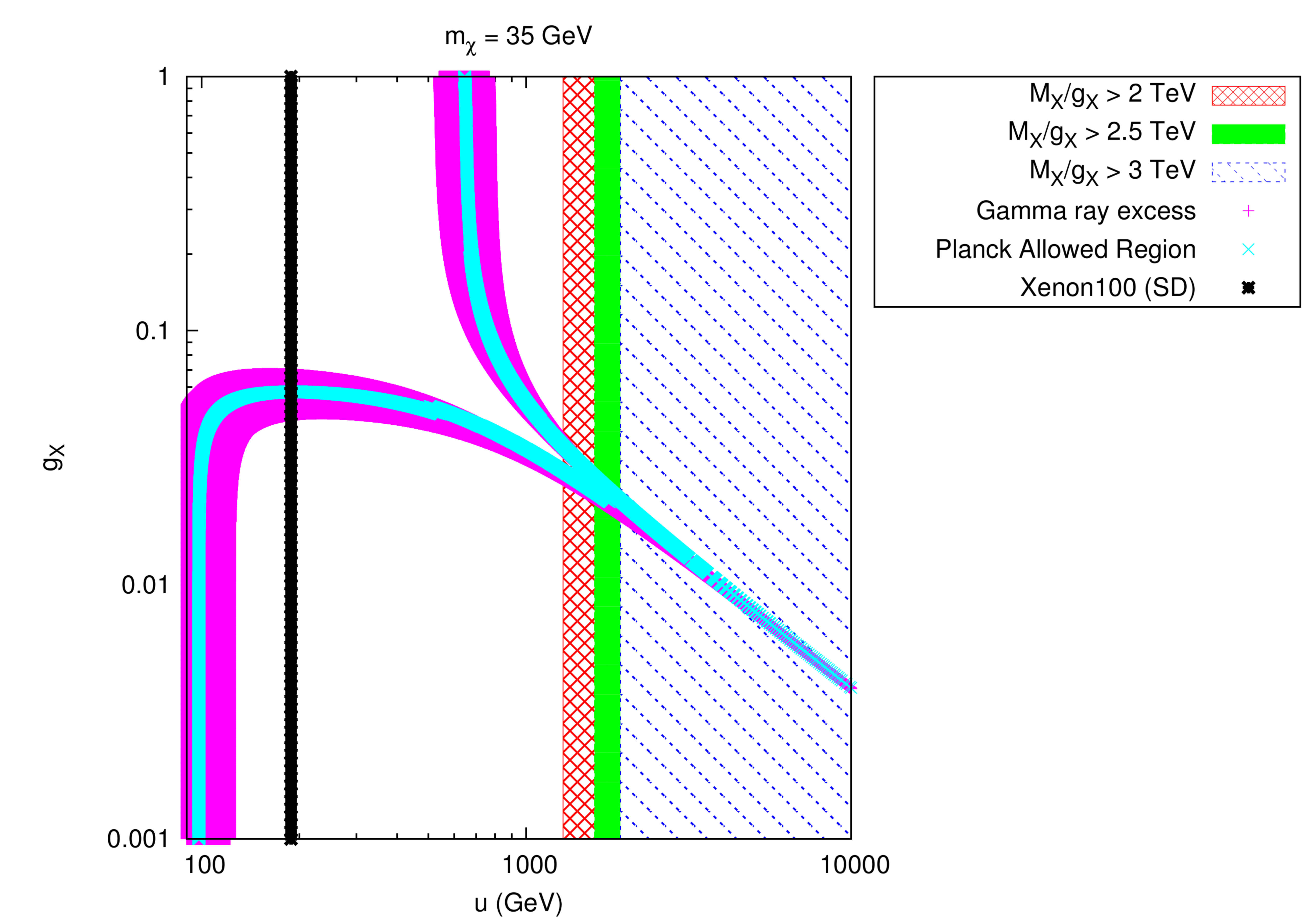,width=1.0\linewidth,clip=}\\
\end{tabular}
\caption{Parameter space in the $g_X-u$ plane for dark matter mass $m_{\chi} = 35$ GeV. The red-hatched, green and blue dot-dashed regions correspond to the allowed region after the constraints on $M_X/g_X$ are imposed. The area to the left of the black line is ruled out by Xenon100 bounds on direct detection cross section. The solid blue and pink regions correspond to regions favored by the relic density and galactic center excess respectively.}
\label{fig4}
\end{figure*}

\begin{figure}[h!] 
\centering
\begin{tabular}{c}
\epsfig{file=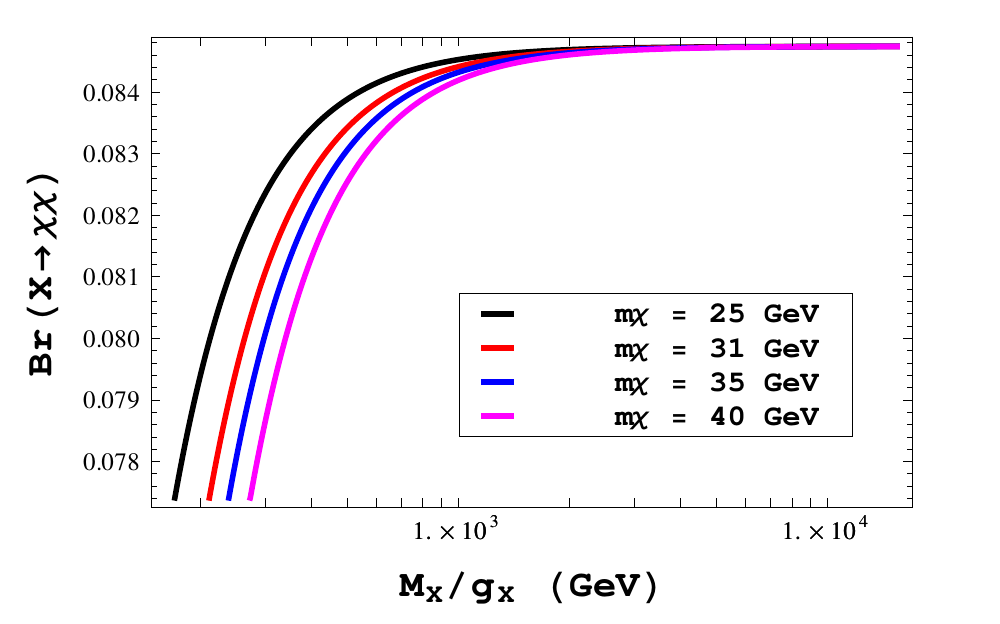,width=1.0\linewidth,clip=}\\
\end{tabular}
\caption{Invisible branching ratio of $X$ boson as a function of $M_X/g_X$}
\label{fig5}
\end{figure}
Using these couplings, we now calculate the dark matter relic abundance for fixed values of dark matter mass and the gauge charges $n_1, n_4$ but with varying $U(1)_X$ gauge coupling $g_X$ and gauge boson mass $M_X$. Similar to our approach in \cite{Borah:2012qr}, here also we make a specific choice of $ n_1$ from which $n_4$ can be found from the normalization $n^2_1 +n^2_4 = 1$. Using the same normalization, the $90\%$ confidence level exclusion on $M_X/g_X$ was shown in \cite{Adhikari:2008uc} where the lowest allowed value of $M_X/g_X$ was found to be approximately $2$ TeV for $\phi = \tan^{-1} (n_4/n_1) = 1.5$. As noted in \cite{Berlin:2014tja}, in order to generate the spectral shape of the gamma ray excess through DM annihilation, the dark matter should be either a $\sim 35 \; 
\text{GeV}$ particle annihilating mostly into $b\bar{b}$ pairs or a $\sim 25 \; \text{GeV}$ particle which annihilates almost democratically to SM fermions. Therefore, we choose these two particular values of dark matter mass in our analysis. After fixing dark matter mass as well as $n_{1,4}$, we vary $g_X$ and $u$ and compute the relic density of dark matter. Instead of assuming a particular value of $x_F$, we first numerically find out the value of $x_F$ which satisfies the following equation
\begin{align}
e^{x_F} - \ln \frac{0.038gm_{PL}m_{\chi}<\sigma v>}{g_*^{1/2}x_F^{1/2}} &= 0
\end{align}
which is nothing but a simplified form of equation (\ref{xf}). For a fixed value of dark matter mass $m_{\chi}$, the annihilation cross section $\sigma$ depends upon $g_X, M_X$. For a particular pair of $g_X$ and $M_X$, we use this value of $x_F$ and compute the relic abundance using equation (\ref{eq:relic}).

The allowed region of parameter space satisfying Planck relic density bound in terms of $g_X, u$ as well as $g_X, M_X$ for $m_{DM} = 25 \; \text{GeV}, 35 \; \text{GeV}$ can be seen in figure \ref{fig1}, \ref{fig2}, \ref{fig3}, \ref{fig4}. We also show the region of parameter space which can give rise to the desired annihilation cross section in order to fit the GC gamma ray excess data. As pointed out by \cite{Berlin:2014tja}, the thermally averaged cross-sections have be  $\langle \sigma v \rangle = (0.77-3.23)\times 10^{-26}\text{cm}^3/s$
and $\langle \sigma v \rangle = (0.63-2.40)\times 10^{-26}\text{cm}^3/s$ for $m_{DM} = 35$ GeV and $m_{DM} = 25$ GeV respectively. From the figures, we see that there are enough overlapping regions of parameter space which can give rise to cross sections required from dark matter relic density constraints as well as from the requirement of giving rise to GC gamma ray excess.

We then take into account of the experimental bounds from dark matter direct detection experiments. Being a Majorana fermion, the dark matter particle in our model gives rise to spin dependent scattering cross section off nuclei mediated by $X$ boson. The latest bound on this cross section is given by the Xenon100 experiment \cite{Aprile:2013doa}. This spin dependent cross section is given by
\begin{align}
\sigma_{SD} &= \frac{4\mu^2_{\chi N}}{\pi M^4_X}g^4_{\chi a}J_N(J_N+1)
\bigg{(}\frac{\langle S_p\rangle}{J_N}(2\Delta^{(p)}_u +\Delta^{(p)}_d) \nonumber \\ 
&+\frac{\langle S_n\rangle}{J_N}(2\Delta^{(n)}_d+\Delta^{(n)}_u)\bigg{)}
\end{align}
where 
$$ \mu_{\chi N} = \frac{m_\chi m_N}{m^2_{\chi}+m^2_N}$$ 
and $J_N$ is the spin of the Xenon nucleus used.
The standard values of the nuclear quark content are taken as$\Delta^{(p)}_u=\Delta^{(n)}_d=0.84$ and $\Delta^{(n)}_u=\Delta^{(p)}_d=-0.43$ \cite{pdg}.
The average spins $\langle S_p\rangle$ and $\langle S_n \rangle$ of the Xenon nucleus are taken from \cite{Aprile:2013doa} as given in table \ref{table1:nuc}.
\begin{table}[!h]
\centering
\begin{tabular}{|c|c|c|}
\hline \hline 
Nucleus & $\langle S_n\rangle$ & $\langle S_p \rangle$ \\
\hline
$^{129}Xe$ & 0.329 & 0.010 \\
$^{131}Xe$ & -0.272 & -0.009 \\
\hline
\end{tabular}
\caption{Average Spin of Nucleus}
\label{table1:nuc}
\end{table}
Xenon100 experiment gives the lowest upper bound on spin dependent cross section as $3.5 \times 10^{-40} \; \text{cm}^2$ for a WIMP mass of 45 GeV at $90\%$ confidence level. Here we take this conservative upper bound for both 25 GeV and 35 GeV dark matter analysis and draw the exclusion line. As can be seen from figure \ref{fig1}, \ref{fig2}, \ref{fig3} and \ref{fig4}, the black solid line corresponds to this direct detection bound such that the parameter space towards the left of this line is ruled out.

To apply the collider bounds on $M_X$ and $g_X$ we follow the analysis of \cite{collider} which studies the scenario of a new heavy abelian gauge boson coupling to dark matter as well as SM fermions in the light of collider and dark matter direct detection data. As discussed by the authors of \cite{collider}, the Large Hadron Collider (LHC) bounds on abelian vector boson coupled to SM, which is approximately $M_X \gtrsim 2.5$ TeV, can be relaxed if $X$ has non-negligible couplings to dark matter. They showed that for $X$ decaying into SM particles with branching ratio $90\%$ and $g_X = 0.1$, the lowest allowed value of $M_X/g_X$ is approximately $2.6$ TeV. This limit gets pushed up to 4 TeV and $4.4$ TeV, if $g_X$ is increased to $0.3$ and weak gauge coupling $g$ respectively. To implement these bounds we compute the branching ratio of $X$ decaying into dark matter particles and plot them as a function of $M_X/g_X$ in figure \ref{fig5}. It can be seen that the maximum branching ratio is around $8.5\%$. We then apply three different lower limits on $M_X/g_X$ namely, 2 TeV, $2.5$ TeV and 3 TeV and check how much of the parameter space remains. These limits can be seen in figure \ref{fig1}, \ref{fig2}, \ref{fig3} and \ref{fig4} in three different colors such that, the parameter space towards the left of that region is ruled out. It can be seen that even if we take a conservative bound $M_X/g_X > 3$ TeV, then also we have available parameter space which satisfies all other dark matter constraints. It should also be noted that near the resonance region in the $g_X-M_X$ as well as $g_X-u$ planes, the values of $g_X$ is much below $0.1$ for which the collider bound is $M_X/g_X > 2.6$ TeV as mentioned above. Thus the bound $M_X/g_X$ is supposed to get further relaxed as we go below $g_X \sim 0.1$ resulting in more and more allowed parameter space.

We note that the parameter space shown in the $g_X-M_X$ plane has regions where mass of dark matter is larger than $M_X$ allowing the possibility of dark matter annihilation into two $X$ bosons. This particular case however, corresponds to the region towards the left side of the plot where $M_X < 25, 35$ GeV. In our model, such annihilation of dark matter into two $X$ bosons can occur through t-channel exhange of the Majorana fermion singlet dark matter itself. Since, the allowed parameter space after incorporating all the constraints correspond to the s-wave resonance region where $M_X \approx 2m_{DM}$, such annihilation of dark matter into two $X$ bosons will not alter the allowed parameter space and hence we have not included this process in our calculations.
\begin{figure}[h!] 
\centering
\begin{tabular}{c}
\epsfig{file=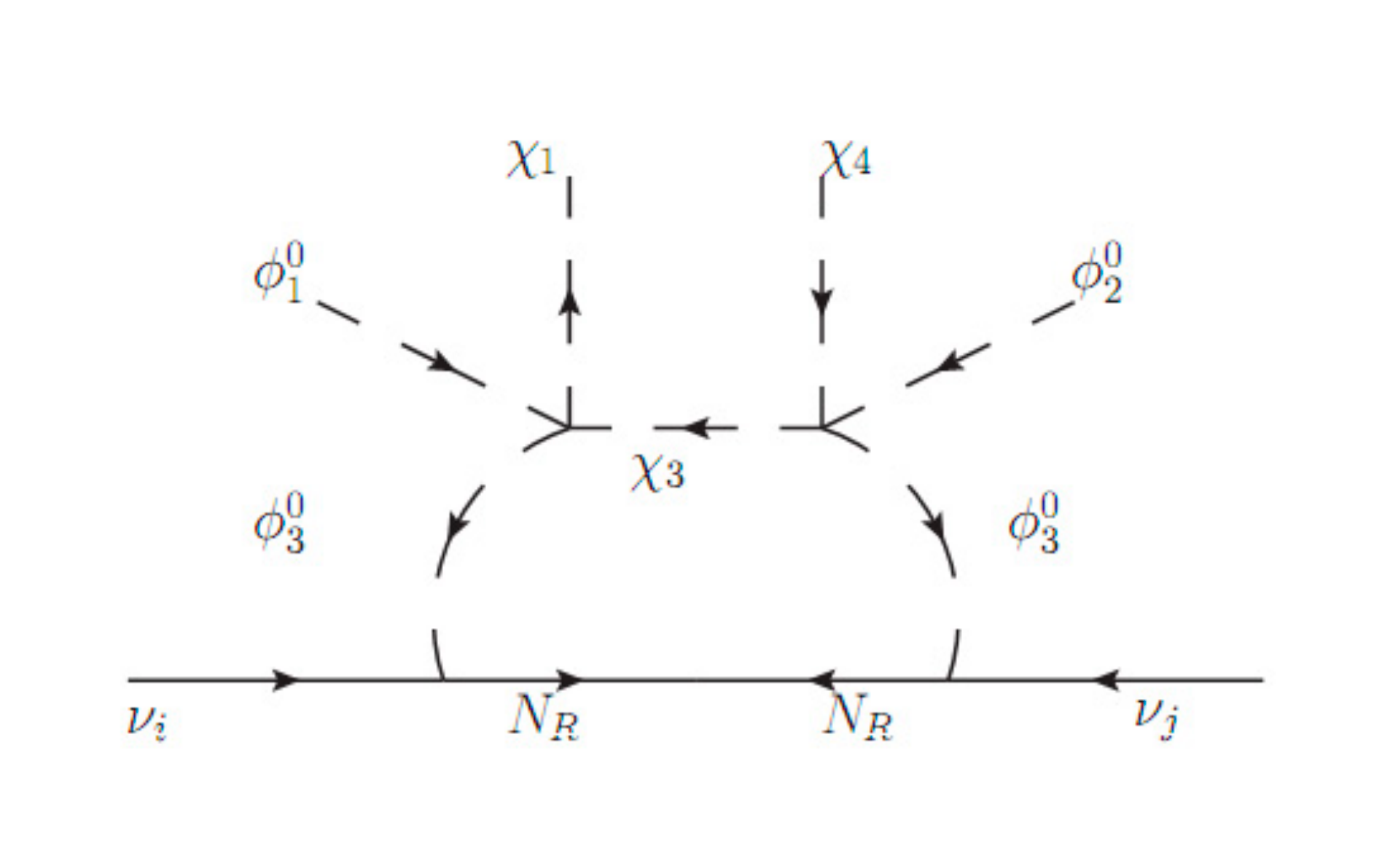,width=0.75\linewidth,clip=}\\
\end{tabular}
\caption{Neutrino mass at one loop level}
\label{fig6}
\end{figure}
\section{Compatibility with Light Neutrino Mass}
\label{neutrino}
As discussed in details in \cite{Borah:2012qr}, the SM light neutrino mass can arise at one loop level in this model as can be seen in figure \ref{fig6}. At tree level only one of the neutrinos acquires non-zero mass from usual type I seesaw mechanism \cite{ti} where the singlet fermion $S_{1R}$ acts as the heavy right handed neutrino. Writing the Yukawa Lagrangian for our model as
$$ \mathcal{L}_Y \supset y \bar{L} \Phi^{\dagger}_1 S_{1R} + h_N \bar{L} \Phi^{\dagger}_3 N_R + h_{\Sigma}  \bar{L}\Phi^{\dagger}_3 \Sigma_R + f_N N_R N_R \chi_4+ f_S S_{1R} S_{1R} \chi_1 $$
\begin{equation}
+ f_{\Sigma} \Sigma_R \Sigma_R \chi_4 + f_{NS} N_R S_{2R} \chi^{\dagger}_2 + f_{12} S_{1R} S_{2R} \chi^{\dagger}_3
\label{yukawa} 
\end{equation}
the tree level light neutrino mass can be written as
\begin{equation}
m_{\nu} \approx \frac{ 2y^2 v_1^2}{f_S u_1}
\label{neutmass}
\end{equation} 
From figure \ref{fig2} and \ref{fig4}, we see that the allowed region from dark matter as well as collider constraints suggest $u_1 = u_2 = u_4 = u \gtrsim 2$ TeV. Since $v_1 \sim 100$ GeV, for neutrino mass to be of eV scale, the equation (\ref{neutmass}) suggest that the Yukawa couplings $y$ have to be fine tuned to $10^{-5}$ which is approximately same as the electron Yukawa coupling in the SM. The other two SM neutrinos can acquire non-zero masses only when loop contributions in figure \ref{fig6} are taken into account. As discussed in \cite{Borah:2012qr}, the  one-loop contribution $(M_\nu)_{ij}$ to neutrino mass is given by
\begin{eqnarray} 
({M_\nu)}_{ij} \approx  \frac{f_3 f_5 v_1 v_2 u_1 u_4}{16 \pi^2} \sum_k {h_{N, \Sigma} }_{ik} {h_{N, \Sigma} }_{jk} \left( A_k +{(B_k)}_{ij} \right)
\label{nuradmass}
\end{eqnarray}
Assuming all the scalar masses in the loop diagram to be almost degenerate and written as $m_{sc}$ then 
\begin{eqnarray}
A_k + (B_k)_{ij} \approx m_{2k} \left[\frac{m_{sc}^2 
+ m_{2k}^2 }{m_{sc}^2 \left( m_{sc}^2 - m_{2k}^2 \right)^2 }- \frac{(2-\delta_{ij})\; m_{2k}^2}{\left(m_{sc}^2 - m_{2k}^2 \right)^3}\ln \left( m_{sc}^2/m_{2k}^2 \right)   \right],
\label{scaldeg}
\end{eqnarray}
where $(M_{N, \Sigma})_k = m_{2k}$. For fermion singlet light dark matter, $m_{2k} \ll m_{sc}$ and hence the above expression can be approximated as 
$$ A_k + (B_k)_{ij} \approx \frac{m_{2k}}{m^4_{sc}} $$
The one-loop neutrino mass can be written as
\begin{eqnarray} 
({M_\nu)}_{ij} \approx  \frac{f_3 f_5 v_1 v_2 u_1 u_4}{16 \pi^2} \sum_k {h_{N, \Sigma} }_{ik} {h_{N, \Sigma} }_{jk} \left( \frac{m_{2k}}{m^4_{sc}}\right)
\end{eqnarray}
Taking $u_1, u_4, m_{sc}$ to be at few TeV's, $v_1, v_2$ at electroweak scale and the singlet mass $m_{2k}$ at few tens of GeV (for light fermion singlet dark matter), the above expression can give rise to eV scale neutrino mass if 
$$ f_3 f_5 h_N h_N \sim 10^{-7}$$
which can be achieved if each of the dimensionless couplings is tuned to be around $10^{-2}$. Thus, a light fermion singlet dark matter of mass 25 GeV or 35 GeV is consistent with the requirement of eV scale SM neutrino masses.

\section{Results and Conclusion}
\label{conclude}
We have studied fermion singlet dark matter in the light of recently observed galactic center gamma ray excess within the framework of an abelian extension of standard model. The model not only gives rise to a stable dark matter candidate, but also gives rise to tiny neutrino masses both at tree level as well as one-loop level. We take two different dark matter masses $m_{\chi} = 25 \; \text{GeV}, 35\; \text{GeV}$ and check whether they can give rise to the desired annihilation cross section in order to satisfy dark matter relic density constraint as well as annihilation into $b\bar{b}$ pairs to explain the galactic center gamma ray excess. We also take into account the constraints from dark matter direct detection experiments on spin dependent scattering cross section of dark matter off nuclei. Since the annihilation and scattering of light fermion singlet dark matter is mediated by the abelian vector boson $X$, these scenarios can also be constrained from LHC limits on additional gauge boson mass $M_X$ and its coupling $g_X$. Without performing a detailed calculation for collider signatures, we use the results from \cite{collider} where the authors found the lower bound on $M_X/g_X$ to be $2.6$ TeV for $\text{BR}(X \rightarrow \text{SM}) = 90\%$ and $g_X = 0.1$. In the present work, we find the maximum branching ratio of $X$ boson into SM particles to be approximately $90\%$. We find that, even after applying a conservative lower limit on $M_X/g_X$ as $3$ TeV, we still have parameter space which can satisfy all dark matter constraints. The allowed parameter space is essentially an s-wave resonance region where mass of the $X$ boson is twice that of dark matter mass. Since the allowed region of parameter space is limited, these scenarios can be further constrained or even ruled out by future data from dark matter direct detection as well as collider experiments.

\begin{acknowledgments}
DB would like to thank the organizers of the workshop "Frontiers of Physics: Colliders and Beyond" during 23-27 June, 2014 at ICTP, Trieste where this work was started. AD likes to thank Council of Scientific and Industrial Research, Govt. of India for financial support through Senior Research Fellowship
(EMR No. 09/466(0125)/2010-EMR-I).
\end{acknowledgments}

\bibliographystyle{apsrev}

\end{document}